\documentstyle[prl,aps,preprint]{revtex}

\begin{document}
\title{Remarks on $2+1$ Self-dual Chern-Simons Gravity}
\author{H. Garc\'{\i}a-Compe\'an$^{a}$\thanks{%
E-mail: compean@fis.cinvestav.mx}, O. Obreg\'on$^{b}$\thanks{%
E-mail: octavio@ifug3.ugto.mx}, C. Ram\'{\i}rez$^c$\thanks{%
E-mail: cramirez@fcfm.buap.mx} and M. Sabido$^{b}$\thanks{%
E-mail: msabido@ifug3.ugto.mx}}
\address{$^{a}$ {\it Departamento de F\'{\i}sica, Centro de Investigaci\'on y de}\\
Estudios Avanzados del IPN\\
P.O. Box 14-740, 07000, M\'exico D.F., M\'exico\\
$^b$ {\it Instituto de F\'{\i}sica de la Universidad de Guanajuato}\\
P.O. Box E-143, 37150, Le\'on Gto., M\'exico\\
$^c$ {\it Facultad de Ciencias F\'{\i}sico Matem\'aticas, Universidad}\\
Aut\'onoma de Puebla\\
P.O. Box 1364, 72000, Puebla, M\'exico}
\date{\today}
\maketitle

\begin{abstract}
\vskip-1.4truecm 
We study $2+1$ Chern-Simons gravity at the classical action
level. In particular we rederive the linear combinations of the ``standard''
and ``exotic'' Einstein actions, from the (anti) self-duality of the
``internal'' Lorentzian indices. The relation to a genuine four-dimensional
(anti)self-dual topological theory greatly facilitates the analysis and its
relation to hyperbolic three-dimensional geometry. Finally a non-abelian
vector field ``dual'' action is also obtained.
\end{abstract}

\vskip -.5truecm





\newpage

\section{Introduction}

\setcounter{equation}{0}

$2+1$ general relativity has many remarkable features such as
renormalizability, integrability, topological invariance etc., which have
been of great value as an heuristic guide in the quest of the correct theory
of the quantum gravitational field in four dimensions (for a review, see 
\cite{carlip}).

These properties derive from the fact that $2+1$ Einstein gravity can be
written as a Chern-Simons gauge theory. This formulation is deeply related
to the celebrated BTZ black hole solution \cite{btz} and recently this
relation has been of great importance in the further development of the
AdS/CFT correspondence \cite{ads}, in particular in the AdS$_3$/CFT$_2$
correspondence in the context of string theory \cite{ads3}. In the AdS$_3$
case, in the context of topological field theories, this correspondence
can be also realized as the well-known correspondence between a $2+1$
Chern-Simons gravity theory on the bulk three-manifold $X$ and the
rational conformal field theory WZW$_2$ on the boundary of this bulk (that
is the CS$_3$/WZW$_2$ correspondence) \cite{wittencs}. 

In Refs. \cite{achu,wittenone} it was found that general relativity in
$2+1$ dimensions with cosmological constant $\lambda $ is equivalent to a
pure Chern-Simons gauge theory on an arbitrary three-manifold $X$ and
certain gauge group ${\bf G}$. $X$ is locally homogeneous with curvature
equal to $\lambda $ and for definiteness we will take it to be given by
$\Sigma \times {\bf R}$ with $\Sigma $ a Riemann surface. For $\lambda
>0$, the gauge group ${\bf G}$ is the de Sitter group SO$(3,1)$; for
$\lambda <0$, it is the anti-de Sitter group SO$(2,2)$ and for $\lambda
=0$, it is the Poincar\'{e} group ISO$(2,1)$. In Ref. \cite{wittenone}
Witten showed, that there are two actions associated to two different ways
of parametrizing the invariant quadratic form on the Lie algebra of ${\bf
G}$; namely, the ``standard'' Einstein action which characterize
classically gravity in $2+1$ dimensions and the ``exotic'' one. It is
interesting to note that each of these actions gives the same classical
equations, but their quantization is expected to lead to two different
quantum theories.  Topology change of space-time (with $\lambda = 0$) was
subsequently discussed within this context \cite{wittentwo}. Some further
developments of Chern-Simons supergravity and extended supergravity were
considered in \cite{vaz}. Applications of Chern-Simons gravity and
supergravity to the two particle scattering problem were studied in
\cite{vaztwo}. 

In this paper we will follow the philosophy of MacDowell and Mansouri (MM)
\cite{mm} and apply it to the 2+1 Chern-Simons gravity. Within this
philosophy, the Chern-Simons action depends only on the gauge field of the
group under consideration. With the usual diagonal Cartan group metric and
the identification of its components with the spin connection and the
vierbein we obtain the ``exotic'' action .

This $2+1$-MM type of action is then naturally generalized by taking its
(anti)self-dual part with respect to the internal indices of the gauge
field. The procedure followed is similar to that in a previous work \cite
{nso}, in which the $3+1$-MM action was generalized for the (anti)self-dual
gauge field. It was shown there that this model is a generalization of the
Pleba\'{n}ski-Ashtekar dynamical action \cite{pleba}. We show
that the two different classical actions, the ``standard''
and the ``exotic'' ones, describing Chern-Simons gravity, are encoded in the
(anti)self-duality of the internal indices of the structure group ${\bf G}$
of the frame bundle. This (anti)self-dual theory corresponds, in 3+1
dimensions, to both the MM self-dual action \cite{nso} and the
Pleba\'{n}ski-Ashtekar formulation \cite{pleba}.

Linear combinations of the ``exotic'' and the ``standard'' actions have been
already considered in the literature \cite{achu,wittenone,adst,lee,ash}.
However, for this pure topological $2+1$ Chern-Simons theory, \ these two
actions \ appear together by considering the (anti)self-dual action. It has
been previously shown, that for a pure $3+1$ topological gravitational model 
\cite{towards,pursuing}, the same kind of procedure gives the sum of the
Euler \ characteristic $\chi (X)$ plus $i\sigma (X)$, with $\sigma (X)$
being the signature, in a way the Euler characteristic seems to play the
role of a ``$\theta $-term'' \cite{wittenthree} (Being both terms
topological one could also say that $\sigma (X)$ is a ``$\theta $-term'').
Similarly, here the ``standard'' (or ``exotic'') action will play the role
of a ``$\theta $-term''. It is well known that the precise relation between
the classical topological invariants in four dimensional manifolds $W$ and
the Chern-Simons invariant $CS$ in its non-trivial boundary $\partial W=X$
is through the Atiyah-Patodi-Singer formula \cite{aps}

\begin{equation}
\sigma (W)=\int_{W}P_{1}(W)-\int_{X}CS-\frac{1}{2}\eta _{S},  \label{patodi}
\end{equation}
where $P_{1}(W)$ is the first Pontrjagin class of $W$ {\it i.e.} $
\int_{W}Tr(R\wedge R)$, $CS$ is the Chern-Simons action and $\eta _{S}$ is
the $\eta $-invariant of $X$. The restriction to self-dual (or
antiself-dual) part of the curvature in $W$, with respect to the group
indices, leads to the Atiyah-Patodi-Singer formula of the form

\begin{equation}
\sigma (W^{\pm })=\int_{W^{\pm }}P_{1}(W^{\pm })-\int_{X}CS^{\pm }-\frac{1}{2
}\eta _{S}^{\pm },  \label{patodiself}
\end{equation}
where $W^{\pm }$ denotes the space $W$ with self-dual (or antiself-dual)
curvature tensor. We will find in Sec. III that keeping this notion of
(anti)self-duality some set of known data find a better understanding with
these data correlated by this (anti)self-duality.

It is known from \cite{wittenone} that some quantum aspects of Chern-Simons
gravity are related to the mathematical description of hyperbolic geometry
of the three-manifold $X$ \cite{thurston,meyerhoff,neumann}. In fact it was
shown in \cite{wittenone} that the volume and Chern-Simons invariants $V$
and $CS$, are related to the ``standard'' and ``exotic'' actions
respectively. However in order to distinguish hyperbolic three-manifolds $X$
with the same invariants $V$ and $CS$, a more refined invariant is needed,
the $\eta $-invariant \cite{yoshida,meyerhofftwo}. In this work we
exhibit that the $\eta $-invariant is originated, in the Chern-Simons
gravity theory, from its relation to an (anti)self-dual topological gravity
theory in four dimensions. Thus all invariants $V$, $CS$ and $\eta_S$ are
encoded in a self-dual topological gravity theory in four dimensions.

In previous works \cite{towards,pursuing}, we were able to exhibit ``field
theory duality'' actions, where the dual action appears with inverted
coupling constants (for a review of duality in field and string
theory, see for example \cite{quevedo}). This was performed for a pure topological
theory and for $3+1$ MM-gravity and supergravity. For Chern-Simons gravity,
a ``field theory duality'' will be found in this paper, in the sense of
Rocek and Verlinde \cite{oganor}, which also gives for the dual action a non
linear sigma-model of the Freedman-Townsend type\cite{freed}, that in this
case can be further integrated to a Chern-Simons action with inverted
coupling constant.

This work is organized as follows: In section II we give an overview of 2+1
Chern-Simons gravity. In section III the consideration of the
(anti)self-dual spin-connection (with respect to the internal ``Lorentzian''
indices) leads to the ``standard'' and ``exotic'' terms, which emerge as a
linear combination. Also in this same section the correspondence with some
aspects of hyperbolic geometry is described. A ``dual'' theory to
Chern-Simons gravity is presented in section IV. In section V we finally
give our concluding remarks.

\vskip2truecm 

\section{Overview of $2+1$ Chern-Simons Gravity}

In this section we will briefly recall the basic structure of $2+1$
Chern-Simons gravity and the relevant structure of the fields, Lagrangians
and symmetries which we will need in the following sections. For a more
complete treatment see Refs. \cite{carlip,wittenone}. In particular we will
follow the notation of these references.

In \cite{achu,wittenone} it was shown that the $2+1$ Einstein-Hilbert action
with vanishing cosmological constant is equivalent to a Chern-Simons action
in $2+1$ dimensions with gauge group given by the Poincar\'{e} group
ISO(2,1).

In the case of nonvanishing cosmological constant $\lambda \not= 0$, Witten
found that the natural generalization is given by the ``standard'' Einstein
action \cite{wittenone}

\begin{equation}
I=\int_{X}\varepsilon ^{ijk}\bigg(e_{ia}(\partial _{j}\omega
_{k}^{a}-\partial _{k}\omega _{j}^{a})+\varepsilon _{abc}e_{i}^{a}\omega
_{j}^{b}\omega _{k}^{c}+{\frac{1}{3}}\lambda \varepsilon
_{abc}e_{i}^{a}e_{j}^{b}e_{k}^{c}\bigg).  \label{standard}
\end{equation}
This Einstein-Hilbert action gives rise to two spacetimes, one for $\lambda
>0$ whose covering is a portion of the de Sitter space with symmetry group
SO(3,1) and the other one with $\lambda <0$ with symmetry group being
SO(2,2). There is another general action, which can be constructed by taking
the standard diagonal representation of the invariant quadratic form of the
Lie algebra of the gauge group \cite{wittenone}. This action was termed the
``exotic'' action and it is given by

\begin{equation}
\tilde{I}=\int_{X}\varepsilon ^{ijk}\bigg(\omega _{i}^{a}(\partial
_{j}\omega _{k}^{a}-\partial _{k}\omega _{j}^{a}+{\frac{2}{3}}\varepsilon
_{abc}\omega _{j}^{b}\omega _{k}^{c})+\lambda e_{i}^{a}(\partial
_{j}e_{k}^{a}-\partial _{k}e_{j}^{a})+2\lambda \varepsilon _{abc}\omega
_{i}^{a}e_{j}^{b}e_{k}^{c}\bigg).  \label{exotic}
\end{equation}
This action has the same classical symmetries as the ``standard'' one and
for this reason it can be added to the ``standard'' Einstein action. Classically, both ``standard'' and ``exotic'' actions are equivalent. In a
representation of the Lie algebra where its generators are given by $
(J_{a}^{+},J_{a}^{-})$ where $J_{a}^{\pm }={\frac{1}{2}}(J_{a}\pm {\frac{1}{
\sqrt{\lambda }}}P_{a})$ for $\lambda \not=0$, the connections are expressed
as ${^{\pm }}A_{i}={^{+}}A^{a}J_{a}^{+}\pm {^{-}}A^{a}J_{a}^{-}$ where ${
^{\pm }}A_{i}^{a}=\omega _{i}^{a}\pm \sqrt{\lambda }e_{i}^{a}$. For $\lambda
<0$, the group SO$(2,2)$ is isomorphic to $SL(2,{\bf R})\times SL(2,{\bf 
R})$ and 
consequently it undergoes a splitting over the real numbers. For $\lambda
>0$, the Lie group SO$(3,1)$ does split as well when complexified, 
to $SL(2,{\bf C} )\times SL(2,{\bf C})$.

The Chern-Simons action reads then
\begin{equation}
I{^{\pm }}=\int_{X}\varepsilon ^{ijk}\big(2{^{\pm }}A_{i}^{a}\partial _{j}{
^{\pm }}A_{k}^{a}+{\frac{2}{3}}\varepsilon _{abc}{^{\pm }}A_{i}^{a}{^{\pm }}
A_{j}^{b}{^{\pm }}A_{k}^{c}\big).
\end{equation}
In terms of the $I^{\pm }$ the ``standard'' and ``exotic'' actions are
written respectively as

\begin{equation}
I={\frac{1}{2\sqrt{\lambda }}}(I^{+}-I^{-}),\ \ \ \ \ \ \ \tilde{I}={\frac{1
}{2}}(I^{+}+I^{-}).  \label{actions}
\end{equation}

Under an euclidean continuation to go from Minkowski space-time to the
euclidean one, the ``standard'' term is real and the only possible
modification (for renormalizing purposes) is a rescaling of the vierbein by
the constant $\hbar $. Moreover, the ``exotic'' term should arise as pure
imaginary and the parameter $k$ in its coefficient should be quantized.
Finally it is worth to mention that one can define the euclidean partition
function

\begin{equation}
Z(X)=\int {\cal D}e{\cal D}\omega e^{-\hat{I}},  \label{pf}
\end{equation}
of the appropriate linear combination of the ``exotic'' and ``standard''
actions to be

\begin{equation}
\hat{I}={\frac{1}{\hbar }}I+i{\frac{k}{8\pi }}\tilde{I},  \label{suma}
\end{equation}
with $k\in {\bf Z}$. In the limit $\hbar \rightarrow 0,$ $Z(X)$ could
``flow'' towards a critical point where the partition function would be described
by two geometric invariants of the hyperbolic geometry of three-manifolds:
the volume $V$ and the Chern-Simons invariant $CS$. These invariants are
described by the ``standard'' and the ``exotic'' actions respectively.

\vskip 2truecm 

\section{The case of the self-dual spin connection}

In this section we will work out the Chern-Simons Lagrangian for (anti)self-dual
gauge connection with respect to duality transformations of the internal
indices of the gauge group ${\bf G}$, in the same philosophy of MM \cite{mm},
and of \cite{nso}. We will show that the two actions ``standard''
and ``exotic'' arise in a natural manner. The action is given by

\begin{equation}
L{^{\pm}}_{CS} = \int_X \varepsilon^{ijk} \bigg( {^{\pm}} A_i^{AB}
\partial_j {\ ^{\pm}} A_{kAB} + {\frac{2 }{3}} {^{\pm}} A_{iA}^B {^{\pm}}
A_{jB}^C {^{\pm}} A_{kC}^A \bigg),  \label{csself}
\end{equation}
where $A,B,C,D= 0,1,2,3,$ $\eta_{AB} = diag(-1,+1,+1,+1)$ and the complex (anti) self-dual connections are
\begin{equation}
{^{\pm}} A_{i}^{AB} = {\frac{1}{2}}(A_{i}^{AB} \mp {\frac{i}{2}}
\varepsilon^{AB}_{ \ \ CD} A_{i}^{CD}).
\end{equation}
Which satisfy the relation $\varepsilon^{AB}_{ \ \ CD} {^{\pm}
} A_i^{CD} = \pm i {^{\pm}} A^{AB}.$

We can compute the first term of the right hand side of the action (\ref
{csself}) and this gives

\begin{equation}
\varepsilon^{ijk} {^{\pm}} A_i^{AB} \partial_j {^{\pm}} A_{kAB} =
\varepsilon^{ijk} \big({\frac{1 }{2}} A_i^{AB} \partial_j A_{kAB} \mp {\frac{
i }{2}} \varepsilon^{ABCD} A_{iAB} \partial_j A_{kCD} \big),  \label{uno}
\end{equation}
while the second part gives

\begin{equation}
\varepsilon^{ijk} {^{\pm}} A_{iA}^{ \ \ B} {^{\pm}} A_{jB}^{ \ \ C} {^{\pm}}
A_{kC}^{ \ \ A} = {\frac{1}{2}}\varepsilon^{ijk} \big( A_{iA}^{ \ \ B}
A_{jB}^{ \ \ C} A_{kC}^{ \ \ A} \pm {\frac{i }{2}} \varepsilon^{ABCD}
A_{iA}^{ \ \ E} A_{jEB} A_{kCD} \big).  \label{dos}
\end{equation}
Thus using Eqs. (\ref{uno}) and (\ref{dos}), the action (\ref{csself}) can
be rewritten as

\begin{equation}
L{^{\pm }}_{CS}=\int_{X}{\frac{1}{2}}\varepsilon ^{ijk}\bigg(
A_{i}^{AB}\partial _{j}A_{kAB}+{\frac{2}{3}}A_{iA}^{\ \ B}A_{jB}^{\ \
C}A_{kC}^{\ \ A}\bigg)\mp {\frac{i}{4}}\varepsilon ^{ijk}\varepsilon ^{ABCD}
\bigg(A_{iAB}\partial _{j}A_{kCD}+{\frac{2}{3}}A_{i\text{ }A}^{\
E}A_{jEB}A_{kCD}\bigg).  \label{eq13}
\end{equation}

\bigskip

In this expression the first term is the Chern-Simons action for the gauge
group ${\bf G}$, while the second term appears as its corresponding ``$
\theta $-term''. The same result was obtained in $3+1$ dimensions when we
considered the (anti)self-dual MM action \cite{pursuing}, or the
(anti)self-dual $3+1$ pure topological gravitational action
\cite{towards}.  It is interesting to note  that the ``$\theta $-term''
in these theories arise by means of the gauge group indices, taking the
role that usually the space-time indices have in the abelian
\cite{wittenthree} and non-abelian cases. Normally for gauge theories it
is meaningful to have arbitrary coupling constants for the dynamical and
the $\theta $-terms (as in the $3+1$-d case \cite {pursuing} ), for that
purpose we need a linear combination of the self-dual and antiself-dual
actions (9). This matter is further discussed in the next section. 

One should remark that the two terms in the action (13)  are the
Chern-Simons and the corresponding ``$\theta $-term'' for the gauge group
${\bf G}$ under consideration. After imposing the particular
identification 
$A_i^{AB}=(A_{i}^{ab},A_{i}^{3a})=(\omega_{i}^{ab},\sqrt{\lambda
}e_{i}^{a})$ and $\omega_{i}^{ab}=\varepsilon^{abc} \omega_{ic}$, the
``exotic'' and ``standard'' actions for the gauge group $SO(3,1)$ are
given by

\[
L_{CS}{^{\pm}} = \int_X {\frac{1}{2}} \varepsilon^{ijk} \bigg(
\omega^a_i(\partial_j \omega_{ka} - \partial_k \omega_{ja}) + {\frac{2 }{3}}
\varepsilon_{abc} \omega_i^a \omega_j^b \omega_k^c + \lambda
e^a_i(\partial_j e_{ka} - \partial_k e_{ia}) - 2 \lambda
\varepsilon_{abc}e_i^a e_j^b \omega_k^c\bigg)
\]

\begin{equation}
\pm i \sqrt{\lambda}\varepsilon ^{ijk}\bigg(e_{i}^{a}(\partial _{j}\omega _{ka}-\partial
_{k}\omega _{ja})-\varepsilon _{abc}e_{i}^{a}\omega _{j}^{b}\omega _{k}^{c}+{
\frac{1}{3}}\lambda \varepsilon _{abc}e_{i}^{a}e_{j}^{b}e_{k}^{c}\bigg),
\label{eq14}
\end{equation}
plus surface terms. These actions degenerate at $\lambda=0$, giving the
pure SO(2,1) Chern-Simons action. However, we can take a linear
combination of them $\tau L_{CS}{^{+}}+\bar{\tau} L_{CS}{^{-}}$. In this
way, the $\sqrt{\lambda}$ in front of the ``standard" term in (\ref{eq14}) 
can be absorbed by a redefinition of the imaginary part of $\tau$. Thus,
with this linear combination, the $\lambda=0$ case gives us the Einstein
$2+1$ gravity without cosmological constant, plus a topological
SO$(2,1)$ CS term.

This action evidently can be rewritten in terms of the ``standard'' $I$ 
and the ``exotic'' $\tilde{I}$ actions reviewed at section II, Eq.(\ref
{suma}) as 
\begin{equation}
L_{CS}{^{\pm }}={\frac{1}{2}}\tilde{I}\pm iI.  \label{tres}
\end{equation}
For SO$(2,2)$ the same procedure can be followed. In this case the (anti) 
self-dual connections $A_i^{\pm}$ are real, hence the action (\ref{eq14})
is also real, as can be obtained by setting $ \lambda\rightarrow
-\lambda$. Thus, in (\ref{eq14}), both cases $\lambda >0$ and $\lambda <0
$ are encoded in the same framework, which permits a simple analysis of \
three-dimensional hyperbolic geometry. 

Up to here we have worked with the Minkowski signature. We can perform a
Wick rotation and the (anti)self-dual Chern-Simons action (\ref{tres}) will
be modified by a global factor $i$, so the above action can be written as

\begin{equation}
L_{CS}{^{\pm }}=I\pm i{\frac{1}{2}}\tilde{I}.
\end{equation}
This is precisely the relevant linear combination of the ``standard'' and
``exotic'' actions used to describe the hyperbolic geometry of $X$ in the
euclidean case, through the volume form $V$ and the Chern-Simons invariant 
$CS$ \cite{wittenone,thurston,meyerhoff,neumann}. Thus the euclidean
partition function of the (anti)self-dual Chern-Simons action (\ref{csself})
is itself a topological invariant

\begin{equation}
Z(X)=exp\bigg({\frac{V}{\hbar }}\pm 2\pi ikCS\bigg).  \label{invariants}
\end{equation}
This provides a more refined invariant than simply using the volume
invariant $V$ and it has been proved to be a complex analytic function in
the space of smooth deformations in the space of the hyperbolic structures
on $X$ \cite{meyerhoff,neumann}. However, there is another natural
invariant, as we have seen, which is the $\eta $-invariant which does not
enter in the formula (\ref{invariants}). The $\eta $-invariant is very
important since it permits to distinguish two hyperbolic three-manifolds
with the
same volume $V$ and $CS$ invariants, that is, with the same partition
function $Z(X)$ \cite{yoshida}.

At this point one would be worried about the (anti)self-duality for the
Chern-Simons gauge theory because this concept is well defined only in
four dimensions.
This can be easily solved by noting from Eq. (2) that the (anti)self-dual
Chern-Simons action (9) can be naturally related to the classical
topological invariants (Euler and signature) of the four dimensional spaces $
W^{\pm }$ and the $\eta $-invariant. Thus the (anti)self-duality used in
this section is induced by a genuine (anti)self-duality from the theory in
four dimensions. Moreover the needed $\eta $-invariant can be extracted from
the stationary phase evaluation of (9) following [5].

There is another issue, which can be better understood also with the idea of
the (anti)self-duality in the internal indices. Euclidean and Lorentzian
quantum field theory of the Chern-Simons gravity Lagrangians, $L$ and $\bar{L
}$ differ by a factor $i$ as we have seen above. From the point of view of
the geometry of $X$ the change of the Lagrangians are encoded in a change of
the orientation of $X$. From the pure topological Chern-Simons gravity point
of view both Lagrangians are equivalent because the theory is independent of
the metric of $X$. This is not satisfactory because both actions amount to
very different quantum theories. It finds a nice explanation by using the
(anti)self-duality inherited by the four-dimensional theory on $W$. The
reason is as follows: a change of the orientation of the bulk $W$ i.e. $
e_{0}\wedge e_{1}\wedge e_{2}\wedge e_{3}\rightarrow -e_{0}\wedge
e_{1}\wedge e_{2}\wedge e_{3}$ leads to a change of orientation on the
boundary $e_{0}\wedge e_{1}\wedge e_{2}\rightarrow -e_{0}\wedge e_{1}\wedge
e_{2}$, for fixed $e_{3}$. In four dimensions the (anti)self-dual projection
implies the choice of an orientation on $W$. Thus as we have seen this
choice is preserved on the boundary $X$ of $W$ and the choice of $L$ or $
\bar{L}$ corresponds to the choice of $L_{CS}^{\pm }$ in Eq. (\ref{csself}).

Some other new invariants are indeed necessary in order to describe
different hyperbolic three-manifolds with the three classical invariants
equal \cite{meyerhofftwo}. The description of these invariants from the
Chern-Simons gravity and supergravity point of view, deserves further
research.

\vskip 2truecm 

\section{Chern-Simons Gravity Dual Action}

This section is devoted to show that a ``dual'' action to the Chern-Simons
gravity action can be constructed following \cite{oganor}. It is worth to
mention that for the abelian case a dual action to the Chern-Simons action
that does invert the coupling constant was worked out previously by
Balachandran {\it et al}, in the context of the quantum Hall effect theory 
\cite{bala}.

Before we proceed to show the ``duality'' of the Chern-Simons gravity action
under vector field transformations we would like to describe our motivation
for it. If one substitutes Eq. (\ref{actions}) in the linear combination (
\ref{suma}) we get

\begin{equation}
\hat{I}=\tau ^{+}I^{+}-\tau ^{-}I^{-}  \label{taus}
\end{equation}
with 
\begin{equation}
\tau ^{\pm }={\frac{1}{2\sqrt{\lambda }\hbar }}\pm i{\frac{k}{16\pi }}.
\label{deftaus}
\end{equation}
Equation (\ref{taus}) has the typical form of actions transforming under
modular transformations SL$(2,{\bf Z})$ \cite{wittenthree}. Clearly the
first term in Eq. (\ref{deftaus}) is the coupling constant of the gravity
theory (\ref{actions}) while the second term is a kind of $\theta $-term.
Thus one can ask whether there exists some modular transformation which
transforms the partition function (\ref{pf})

\begin{equation}
Z({\frac{a\tau + b}{c\tau +d}}) = (c\tau + d)^u (c\bar{\tau} + d)^v Z(\tau),
\end{equation}
where $u$ and $v$ are the weights of the modular transformation. This would
have importance in the construction of further new invariants in the
description of hyperbolic geometry as we have seen at the end of section
III. In this section we will show that such a transformation exists but
action (\ref{taus}) is in fact self-dual under this SL$(2,{\bf Z})$ group.

We begin from the original non-abelian Chern-Simons action given by

\begin{equation}
L=\int_{X}{\frac{g}{4\pi }}{\rm Tr}(A\wedge H),
\end{equation}
where $H=dA+{\frac{2}{3}}A\wedge A$ and ${\rm Tr}$ is a quadratic form in
the Lie algebra of the relevant gauge group ${\bf G}$. In what follows we
will use the usual conventions: $A=A_{i}^{AB}M_{AB}dx^{i}$, $
G=G_{ij}^{AB}M_{AB}dx^{i}\wedge dx^{j}$, where $M_{AB}$ are the generators
of the Lie algebra of ${\bf G}$ and they satisfy $[M_{AB},M_{CD}]=if_{\ \
ABCD}^{EF}M_{EF}$ with $f_{\ \ ABCD}^{EF}$ the structure constants of the
Lie algebra. The invariant quadratic form on the Lie algebra is given by $
{\rm Tr}(M_{AB}M_{CD})=\eta _{(AC}\eta _{BD)}$. With these conventions, the
above action is written as

\begin{equation}
L=\int_{X}d^{3}x{\frac{g}{4\pi }}\varepsilon ^{ijk}A_{i}^{AB}\bigg(\partial
_{j}A_{kAB}+{\frac{1}{3}}f_{ABCDEF}A_{j}^{CD}A_{k}^{EF}\bigg).  \label{cs}
\end{equation}

Now, as usual we propose a parent action in order to derive the dual action
to (\ref{cs}),

\begin{equation}
L_{D}=\int_{X}a{\rm Tr}(B\wedge H)+b{\rm Tr}(A\wedge G)+c{\rm Tr}(B\wedge G),
\label{parentaction}
\end{equation}
where as usual, $B$ and $G$ are Lie algebra-valued one and two forms over $X$
and $a,b,c$ are constants to be determined. In local coordinates on $X$, the
above action can be written as

\begin{equation}
L_D = \int_X d^3 x \varepsilon^{ijk} \bigg( a B^{AB}_i H_{jkAB} + b A^{AB}_i
G_{jkAB} + c B^{AB}_i G_{jkAB} \bigg),  \label{parent}
\end{equation}
where 
\begin{equation}
H_{jkAB}= \partial_j A_{kAB} + {\frac{1 }{3}} f_{ABCDEF}A^{CD}_jA^{EF}_k.
\label{ache}
\end{equation}
It is a very easy matter to show that Eq. (\ref{cs}) can be derived from the
parent action. To see that, consider first the partition function of the
parent action of the form

\begin{equation}
Z=\int {\cal D}A{\cal D}G{\cal D}Bexp\big(+iL_{D}\big),
\end{equation}

\begin{equation}
exp\bigg(+i\tilde{L}_{D}\bigg)=\int {\cal D}G{\cal D}B\ exp\bigg(
+i\int_{X}d^{3}x\varepsilon
^{ijk}(aB_{i}^{AB}H_{jkAB}+bA_{i}^{AB}G_{jkAB}+cB_{i}^{AB}G_{jkAB})\bigg).
\end{equation}
One can integrate out first with respect $G$. This gives

\begin{equation}
exp \bigg( +i \tilde{L}_D \bigg)=\int {\cal D}B \delta\big(b A^{AB}_i + c
B^{AB}_i \big) exp \bigg(+i a\int_X d^3 x \varepsilon^{ijk} B^{AB}_i
H_{jkAB} \bigg).
\end{equation}
Integration with respect to $B$ gives the final form

\begin{equation}
\tilde{L}_{D}=-{\frac{ab}{c}}\int_{X}d^{3}x\varepsilon
^{ijk}A_{i}^{AB}(\partial _{j}A_{kAB}+{\frac{1}{3}}
f_{ABCDEF}A_{j}^{CD}A_{k}^{EF}).
\end{equation}
A choice of the constants of the form 
\begin{equation}
c=-{\frac{4\pi }{g}},\qquad a=b=1,  \label{valores}
\end{equation}
immediately gives the formula (\ref{cs}).

The ``dual'' action $L^*_D$ can be computed as usually in the euclidean
partition function, by integrating first out with respect to the physical
degrees of freedom $A$

\begin{equation}
exp\bigg( - L^*_D \bigg) = \int {\cal D}A exp \bigg( - L_D \bigg).
\end{equation}
The resulting action is of the gaussian type in the variable $A$ and thus
after some computations it is easy to find the ``dual'' action

\begin{equation}
L_{D}^{\ast }=\int_{X}d^{3}x\varepsilon ^{ijk}\bigg \{-{\frac{3}{4a}}
(a\partial _{i}B_{jAB}+bG_{ijAB})[{\bf R}^{-1}]_{kn}^{ABCD}\varepsilon
^{lmn}(a\partial _{l}B_{mCD}+bG_{lmCD})+c\alpha _{i}^{AB}G_{jkAB}\bigg\},
\label{bosonicdual}
\end{equation}
where $[{\bf R}]$ is given by $[{\bf R}]_{ABCD}^{ij}=\varepsilon ^{ijk}f_{\
\ ABCD}^{EF}B_{kEF}$ whose inverse is defined by $[{\bf R}]_{ABCD}^{ij}[{\bf 
R}^{-1}]_{jk}^{CDEF}=\delta _{k}^{i}\delta _{AB}^{EF}.$

The partition function is finally given by 
\begin{equation}
Z = \int {\cal D}G {\cal D}B \sqrt{det({\bf M}^{-1})} exp \big( - L^*_D
\big).
\end{equation}

In this ``dual action'' the $G$ field is not dynamical and can be
integrated out. If we take the values (\ref{valores}) for the constants of the
parent action (\ref{parentaction}), then the integration of this auxiliary
field gives 
\begin{equation}
Z=\int {\cal D}Bexp\bigg(-{\frac{4\pi }{g}}\varepsilon
^{lmn}(B_{l}^{AB}\partial _{m}B_{nAB}-{\frac{4\pi }{g}}
f_{ABCDEF}B_{l}^{AB}B_{m}^{CD}B_{n}^{EF})\bigg).
\end{equation}

As Balachandran et. al. mention, in the abelian case \cite{bala}, the fields 
$B$ cannot be rescaled if we impose ``periodicity'' conditions on them. Thus,
this dual action has inverted coupling with respect to the original one.

\vskip 2truecm 

\section{Concluding Remarks}

In this paper we have rederived linear combinations of the ``standard''
and ``exotic'' actions given by Witten \cite{wittenone}. An interesting
feature is that both actions are encoded in the (anti)self-duality of the
gauge group ${\bf G}$ of the theory.  Thus both actions are obtained from
the Chern-Simons gravity action for the (anti)self-dual connection
(\ref{csself}) of ${\bf G}$. The usual Chern-Simons action corresponds to
the ``exotic'' one and it's ``$\theta $-term'' gives the ``standard''
action. So to construct a linear combination \ of the Chern-Simons action
\ and it's ``$\theta $-term'' is equivalent to a linear combination \ of
the self-dual and antiself-dual \ actions, as we would have expected. 

As a matter of fact, we have provided the justification of this derivation
by observing that the linear combinations of the actions $I$ and
$\tilde{I}$ come from the (anti)self-duality structure of the four
manifold $W$ when one is restricted to work on the boundary $X$ of $W$. 

Furthermore, for the (non-abelian) Chern-Simons gravity action, we have
found a ``non-abelian dual action'' in the sense of \cite{oganor}. This
resulting ``dual action'' consists of a non linear sigma model
\cite{freed} of the Freedman-Townsend type eq. (32). Moreover, after a
further integration, one can reduce this action to a Chern-Simons action
with inverted coupling eq. (34). It is interesting to remark that all
these calculations are performed at the level of the gauge fields $
A_{i}^{AB}$ and its corresponding Chern-Simons action with a ``$\theta $
-term''. It is only when we identify with gravity, that we come to the
``standard'' and ``exotic'' actions and to a metrical theory. 

The results obtained in section III can be easily generalized to
Chern-Simons supergravity. It is tantalizing to find the supersymmetric
extension of the ``dual'' action obtained in section IV\cite
{pursuing,wittenfour}. Furthermore it would be of some interest to apply
the results of this paper to explore further uses of Chern-Simons gravity
to hyperbolic geometry of $X$ and perturbative expansions of Chern-Simons
theory with non-compact groups \cite{dror}. Work in these directions will
be reported elsewhere. 

\vskip 2truecm 
\centerline{\bf Acknowledgments}

It is a pleasure to thank A.P. Balachandran for useful discussions. This
work was supported in part by CONACyT grant 28454E.


\vskip 2truecm 



\end{document}